\documentclass[12pt]{article}
\usepackage[dvips]{graphicx}

\begin{document}

\title{A Trickiness of the High-Temperature Limit \\ for Number Density 
Correlation Functions \\ in Classical Coulomb Fluids}

\author{L. {\v S}amaj$^1$}

\date{\empty}

\maketitle

\begin{abstract}
The Debye-H\"uckel theory describes rigorously the thermal equilibrium 
of classical Coulomb fluids in the high-temperature $\beta\to 0$ regime
($\beta$ denotes the inverse temperature).
It is generally believed that the Debye-H\"uckel theory and
the systematic high-temperature expansion provide an
adequate description also in the region of small 
{\em strictly positive} values of $\beta>0$.
This hypothesis is tested in the present paper on a two-dimensional 
Coulomb gas of pointlike $+/-$ unit charges interacting via a logarithmic
potential which is equivalent to an integrable sine-Gordon field model.
In particular, we apply a form factor method to obtain the exact
asymptotic large-distance behavior of particle correlation functions,
considered in the charge and number density combinations. 
We first determine the general forms of the leading and subleading asymptotic 
terms at strictly positive $\beta>0$ and then evaluate their high-temperature 
$\beta\to 0$ forms.
In the case of the {\em charge} correlation function, the leading asymptotic
term at a strictly positive $\beta>0$ is also the leading one in 
the high-temperature $\beta\to 0$ regime.
On the contrary, the $\beta\to 0$ behavior of the {\em number density} 
correlation function is accompanied by an interference between the first 
two asymptotic terms. 
Consequently, the large-distance behavior of this function exhibits 
a discontinuity when going from strictly positive values of $\beta>0$ 
to the Debye-H\"uckel limit $\beta\to 0$.
This is the crucial conclusion of the paper: the large-distance asymptotics 
and the high-temperature limit do not commute for the density correlation 
function of the two-dimensional Coulomb gas.
\end{abstract}

\medskip


\vfill

\noindent $^1$ 
Institute of Physics, Slovak Academy of Sciences, D\'ubravsk\'a cesta 9, 
\newline 845 11 Bratislava, Slovak Republic; e-mail: fyzimaes@savba.sk

\newpage

\renewcommand{\theequation}{1.\arabic{equation}}
\setcounter{equation}{0}

\section{Introduction}
The object of study of the present paper is the equilibrium statistical
mechanics of infinite (bulk) classical (i.e., non-quantum) Coulomb fluids. 
For the sake of simplicity, we shall restrict ourselves to the
case of a symmetric Coulomb gas, i.e., a neutral system of two species
of particles $\{ j \}$ of opposite unit charges $\{ q_j = \pm 1 \}$,
living in a $\nu$-dimensional space of points ${\bf r}\in R^{\nu}$.
The system is immersed in a homogeneous medium of dielectric constant
$\epsilon=1$. 
The interaction energy of a set of particles $\{ q_j, {\bf r}_j\}$ is given by
\begin{equation} \label{1.1}
E(\{ q_j, {\bf r}_j\}) = \sum_{j<k} q_j q_k 
v(\vert {\bf r}_j - {\bf r}_k\vert) ,
\end{equation}
where the Coulomb potential $v$ is the solution of the Poisson equation
\begin{equation} \label{1.2}
\Delta v({\bf r}) = - s_{\nu} \delta({\bf r})
\end{equation}
with $s_{\nu}$ being the surface area of the $\nu$-dimensional unit sphere.
In particular,
\begin{equation} \label{1.3}
v({\bf r}) = \left\{
\begin{array}{cc}
- \ln(\vert {\bf r}\vert/r_0) & \nu = 2 , \\
& \\
1/\vert {\bf r}\vert & \nu = 3 .
\end{array} \right.
\end{equation}
The free length scale $r_0$, which fixes the zero point of the 2D Coulomb
potential, will be set to unity for simplicity.
The Fourier transform of the Coulomb potential given by 
the $\nu$-dimensional Poisson equation (\ref{1.2}) exhibits the form 
$1/\vert {\bf k}\vert^2$ with the characteristic singularity at
${\bf k}\to {\bf 0}$.
This maintains many generic properties, like screening and the related
sum rules \cite{Martin}, of ``real'' three-dimensional (3D) Coulomb fluids.
For the case of pointlike particles, the singularity of $v({\bf r})$
at the origin ${\bf r}={\bf 0}$ prevents the thermodynamic stability
against the collapse of positive-negative pairs of charges; in 2D for
small enough temperatures, in 3D for any temperature.
In such case the Coulomb potential must be regularized at short
distance, e.g., by a hard-core potential around each particle.

A complete analysis of the bulk Coulomb gas can be done within 
the framework of the mean-field theory of Debye and H\"uckel (DH) 
\cite{Debye}, sometimes called the linear Poisson-Boltzmann (PB) theory.
The Debye-H\"uckel theory describes rigorously the distribution
functions of the {\em internal} Coulomb-gas charges in 
the high-temperature regime $\beta\to 0$ ($\beta$ is the dimensionless 
inverse temperature) \cite{Kennedy83}. 
The nonlinear (``classical'' in field theory) PB theory, which arises 
as the zeroth-order term in a loop expansion of the grand partition function 
\cite{Netz}, describes rigorously the density profiles of the Coulomb-gas
charges induced by an {\em external} charge distribution, in a specific 
scaling regime of the infinite-temperature limit \cite{Kennedy84}. 
The relation between the linear and nonlinear PB theories is described
in field theoretical books; see, e.g., Ref. \cite{Rivers}.
There exist many phenomenological approximations for finite temperatures
based on heuristic extensions of the mean-field theories \cite{Levin}.

In the rigorous mathematical sense, the high-temperature $\beta\to 0$
regime is not equivalent to the case of the inverse temperature $\beta$
being a sufficiently small, but {\em strictly positive} number, 
say $\beta=10^{-15}$.
On the other hand, it is generally believed that the $\beta\to 0$ 
Debye-H\"uckel theory and its improvement by a systematic $\beta$
expansion adequately describe also the region of small positive values of 
$\beta>0$.
This, at first sight natural, assumption might not be true
for specific statistical quantities of Coulomb fluids and therefore
its validity has to be verified on exact results at strictly positive 
$\beta$.

To solve exactly a 3D Coulomb fluid at strictly positive $\beta$
is a hopeless task.
The situation is more optimistic in the case of 2D logarithmic Coulomb fluids.
The stability of the 2D Coulomb gas of pointlike $\pm 1$ charges against
the collapse, associated with the 2D spatial integrability of 
the Boltzmann factor of the positive-negative pair of charges
$\exp[\beta v({\bf r})] = \vert {\bf r} \vert^{-\beta}$ at short
distances, is restricted to inverse temperatures $\beta<2$.
In this stability region the bulk thermodynamic properties
(free energy, internal energy, specific heat, etc.) of the 2D
Coulomb gas have been obtained exactly based on its equivalence
with the (1+1)-dimensional sine-Gordon theory \cite{Samaj00}.
Later on, the form-factor method was applied to calculate the large-distance
asymptotic behavior of the charge \cite{Samaj02a} and number density 
\cite{Samaj02b} pair correlation functions.
Within the half space geometry, the surface thermodynamics (surface tension)
of the stable 2D Coulomb gas in contact with a grounded ideal conductor wall 
was obtained through its mapping onto the boundary sine-Gordon model
with an integrable Dirichlet boundary condition \cite{Samaj01}. 
The exact nonperturbative asymptotic forms of the charge and number
density profiles of the Coulomb-gas species at large distances
from the conductor wall were calculated in Ref. \cite{Samaj05}.
It was shown that the DH theory adequately describes the charge profile 
at a small strictly positive $\beta$, but this is no longer true for 
the {\em number density} profile.
This surprising result was the primary motivation for the present work.

In this paper, we reconsider the bulk 2D Coulomb gas and perform
a more detailed form-factor analysis of the asymptotic large-distance 
behavior of the charge and number density pair correlation functions.
In Refs. \cite{Samaj02a,Samaj02b}, we have been concerned only with the
exact leading large-distance asymptotics of the correlation functions.
This information is not sufficient in view of present aims for which 
we need also terms which are subleading in the large-distance limit.
We first review briefly technicalities and findings of the
previous works and then develop the form-factor formalism on higher levels
in order to obtain the explicit forms of these subleading terms.
The program is nontrivial and requires the application of some specific 
techniques known in the theory of integrable field models, e.g.,
a bootstrap procedure when calculating the form-factors
of higher (heavier) breathers in the particle spectrum of 
the sine-Gordon theory.
From the technical point of view, the present paper goes far beyond 
the previous ones \cite{Samaj02a,Samaj02b}.

For both the charge and number density correlation functions, we first 
determine the general forms of the leading and subleading asymptotic 
terms at strictly positive $\beta>0$ and then evaluate their 
$\beta\to 0$ forms.
In the case of the {\em charge} correlation function, the leading term at 
positive $\beta>0$ is also the leading one in the $\beta\to 0$ regime.
On the contrary, the $\beta\to 0$ behavior of the {\em number density} 
correlation function is accompanied by an interference between the first 
two asymptotic terms. 
The large-distance behavior of this function therefore exhibits 
a discontinuity when going from strictly positive values of 
$\beta>0$ to the Debye-H\"uckel limit.
Equivalently, the large-distance asymptotics of this function
at fixed temperature $\beta>0$ does not coincide with that obtained
when the high-temperature $\beta$-expansion has been performed first.
This is the crucial conclusion of the paper: the large-distance
asymptotics and the high-temperature limit do not commute for
the density correlation function of the 2D Coulomb gas. 

The paper is organized as follows.
In section 2, we introduce the notation and briefly summarize previous 
results obtained from the mapping of the infinite 2D Coulomb gas onto
the bulk (1+1)-dimensional sine-Gordon theory.
The detailed form-factor analysis of the asymptotic large-distance
behavior of the charge and number density correlation functions 
at a small positive $\beta>0$ is presented in section 3.
The high-temperature $\beta\to 0$ behavior of the obtained results 
is discussed in section 4.
A comparison is made with the systematic high-temperature 
expansion, summarized in the Appendix. 
A recapitulation and some concluding remarks are given in section 5.

\renewcommand{\theequation}{2.\arabic{equation}}
\setcounter{equation}{0}

\section{Sine-Gordon representation}
The bulk 2D Coulomb gas of pointlike $\pm 1$ charges is treated
in the grand canonical ensemble characterized by the inverse temperature
$\beta$ and by the couple of particle fugacities $z_+ = z_- = z$
(at some places, in order to distinguish between the $+$ and $-$ charges,
we shall keep the notation $z_{\pm}$).
The grand partition function is defined by
\begin{equation} \label{2.1}
\Xi = \sum_{N_+,N_-=0}^{\infty} \frac{z_+^{N_+}}{N_+!} 
\frac{z_-^{N_-}}{N_-!} Q(N_+,N_-) ,
\end{equation}
where
\begin{equation} \label{2.2}
Q(N_+,N_-) = \int_{{\rm R}^2} \prod_{j=1}^N d^2 r_j\, 
\exp\left[ - \beta E(\{ q_j,{\bf r}_j\}) \right]
\end{equation}
is the configuration integral of $N_+$ positive and $N_-$ negative charges,
$N = N_+ + N_-$ and the interaction energy $E$ is defined by (\ref{1.1})
with $v$ being the logarithmic Coulomb potential.
The infinite system is homogeneous and translationally invariant.
Denoting the thermal average as $\langle \cdots \rangle_{\beta}$,
the number density of particles of one charge sign $q (= \pm 1)$
is defined by
\begin{equation} \label{2.3}
n_q = \left\langle \sum_j \delta_{q,q_j} \delta({\bf r}-{\bf r}_j)
\right\rangle_{\beta} .
\end{equation} 
Due to the charge symmetry, $n_+ = n_- = n/2$ where $n$ is the total
number density of particles.
At the two-particle level, one introduces the two-body density
\begin{equation} \label{2.4}
n_{qq'}(\vert {\bf r}-{\bf r}'\vert) = 
\left\langle \sum_{j\ne k} \delta_{q,q_j} \delta({\bf r}-{\bf r}_j)
\delta_{q',q_k} \delta({\bf r}'-{\bf r}_k) \right\rangle_{\beta} ,
\end{equation} 
which possesses the obvious symmetry $n_{++} = n_{--}$ and $n_{+-} = n_{-+}$.
We consider also the pair correlation function
\begin{equation} \label{2.5}
h_{qq'}(\vert {\bf r}-{\bf r}'\vert) = 
\frac{n_{qq'}(\vert {\bf r}-{\bf r}'\vert)}{n_q n_{q'}} - 1 ,
\end{equation}
in the charge (subscript $\rho$) and number density (subscript $n$) 
combinations:
\begin{equation} \label{2.6}
h_{\rho} = \frac{1}{4} \sum_{q,q'=\pm 1} q q' h_{qq'} , \quad
h_n = \frac{1}{4} \sum_{q,q'=\pm 1} h_{qq'} .
\end{equation}

The 2D Coulomb gas is mappable onto the sine-Gordon theory \cite{Minnhagen}.
Using the fact that according to Eq. (\ref{1.2}) $-\Delta/(2\pi)$ is 
the inverse operator of the Coulomb potential $v$ and renormalizing 
the particle fugacity $z$ by the (divergent) self-energy term 
$\exp[\beta v(0)/2]$, the grand partition function
can be turned via the Hubbard-Stratonovich transformation into
\begin{equation} \label{2.7}
\Xi(z) = \frac{\int {\cal D}\phi\, \exp[-S(z)]}{\int {\cal D}\phi\, 
\exp[-S(0)]} ,
\end{equation} 
where
\begin{equation} \label{2.8}
S(z) = \int_{{\rm R}^2} d^2 r\, \left[ \frac{1}{16\pi} 
\left( \nabla\phi \right)^2 - 2 z \cos(b\phi) \right] , 
\quad b^2 = \frac{\beta}{4} 
\end{equation}
is the 2D Euclidean action of the (1+1)-dimensional sine-Gordon theory.
Here, $\phi({\bf r})$ is a real scalar field and $\int {\cal D}\phi$
denotes the functional integration over this field.
The one- and two-body densities of the Coulomb gas are expressible
as averages over the sine-Gordon action (\ref{2.8}) as follows
\begin{eqnarray}
n_q & = & z_q \left\langle {\rm e}^{{\rm i}q b \phi} \right\rangle ,
\label{2.9} \\ 
n_{qq'}(\vert {\bf r}-{\bf r}'\vert) & = & z_q z_{q'}
\left\langle {\rm e}^{{\rm i}q b \phi({\bf r})} 
{\rm e}^{{\rm i}q' b \phi({\bf r}')} \right\rangle .
\label{2.10} 
\end{eqnarray}
The renormalized fugacity parameter $z$ gets a precise meaning 
under the short-distance conformal normalization
\begin{equation} \label{2.11}
\left\langle {\rm e}^{{\rm i} b \phi({\bf r})} 
{\rm e}^{-{\rm i}b \phi({\bf r}')} \right\rangle 
\sim \vert {\bf r}-{\bf r}' \vert^{-4 b^2} \quad
\mbox{as $\vert {\bf r}-{\bf r}' \vert\to 0$} ;
\end{equation}
for an explanation, see e.g. Ref. \cite{Samaj03}.

The (1+1)-dimensional sine-Gordon model is an integrable field theory
\cite{Zamolodchikov79}.
Its particle spectrum is the following.
The basic particles are the soliton $S$ and the antisoliton ${\bar S}$
which form a particle-antiparticle pair of equal masses $M$;
for topological reason, the soliton and the antisoliton coexist in pairs.
The $S$-${\bar S}$ pair can create bound states, the so-called
``breathers'' $\{ B_j; j=1,2,\ldots < p^{-1} \}$. 
Their number depends on the inverse of the temperature parameter
\begin{equation} \label{2.12}
p = \frac{b^2}{1-b^2} \left( = \frac{\beta}{4-\beta} \right) .
\end{equation}
The mass of the $B_j$ breather is given by
\begin{equation} \label{2.13}
m_j = 2 M \sin\left( \frac{\pi p}{2} j \right) ,
\end{equation}
and this breather disappears from the particle spectrum just when
$m_j = 2M$, i.e. at the point $p=1/j$.
Note that the breathers exist only in the stability region of
the corresponding Coulomb gas: the lightest $B_1$ breather disappears 
just at the collapse border $p=1$ ($\beta=2$), the $B_2$ breather
at $p=1/2$ ($\beta=4/3$), the $B_3$ breather at $p=1/3$ ($\beta=1$), etc.

The explicit results for the ground-state characteristics of
the (1+1)-dimensional sine-Gordon model in the stability region
were derived quite recently due to a progress 
in the method of Thermodynamic Bethe Ansatz.
The (dimensionless) specific grand potential $\omega$,
defined in the infinite-volume limit as
\begin{equation} \label{2.14}
- \omega = \frac{1}{\vert {\rm R}^2\vert} \ln \Xi ,
\end{equation}
was found by Destri and de Vega \cite{Destri} in the form 
\begin{equation} \label{2.15}
- \omega = \frac{m_1^2}{8\sin(\pi p)} .
\end{equation}
Here, $m_1$ is the mass of the lightest $B_1$ breather [see formula
(\ref{2.13}) taken with $j=1$].
Under the conformal normalization (\ref{2.11}), the relationship between
the soliton mass $M$ and the fugacity $z$ was established 
in Ref. \cite{Zamolodchikov95}:
\begin{equation} \label{2.16}
z = \frac{\Gamma(b^2)}{\pi\Gamma(1-b^2)}
\left[ M \frac{\sqrt{\pi} \Gamma[(1+p)/2]}{2\Gamma(p/2)}
\right]^{2-2b^2} ,
\end{equation}
where $\Gamma$ stands for the Gamma function.
The total number density $n$ of the Coulomb-gas charges is given
by the standard relation
\begin{equation} \label{2.17}
n = z \frac{\partial(-\omega)}{\partial z} .
\end{equation}
Eqs. (\ref{2.15}) and (\ref{2.16}) thus imply the explicit 
density-fugacity relationship, and consequently the complete bulk
thermodynamics, of the 2D Coulomb gas in the whole stability
region $\beta<2$ \cite{Samaj00}.

Using the thermodynamic formulae (\ref{2.14})-(\ref{2.17}),
the mass of the $B_j$ breather (\ref{2.13})
can be expressed as follows
\begin{equation} \label{2.18}
m_j = \kappa \left[ \frac{\pi p}{2} 
\tan\left( \frac{\pi p}{2} \right) \right]^{-1/2}
\sin\left( \frac{\pi p}{2} j \right) ,
\end{equation}
where $\kappa=\sqrt{2\pi\beta n}$ is the 2D inverse Debye length.
The high-temperature $p\to 0$ form of $m_j$ reads
\begin{equation} \label{2.19}
\lim_{p\to 0} m_j = j \kappa \quad (j=1,2,\ldots) .
\end{equation}

\renewcommand{\theequation}{3.\arabic{equation}}
\setcounter{equation}{0}

\section{Form-factor analysis}
In a 2D integrable field theory with spectrum of particles
$\{ \epsilon \}$ of masses $\{ m_{\epsilon} \}$,
the two-point correlation function of local operators ${\cal O}_a$ 
($a$ is a free parameter) can be written as an infinite convergent 
series over multiparticle intermediate states \cite{Smirnov}:
\begin{eqnarray}
\langle {\cal O}_a({\bf r}) {\cal O}_{a'}({\bf r}') \rangle
& = & \sum_{N=0}^{\infty} \frac{1}{N!} \sum_{\epsilon_1,\ldots,
\epsilon_N} \int_{-\infty}^{\infty} \frac{{\rm d}\theta_1 \cdots
{\rm d}\theta_N}{(2\pi)^N} 
F_a(\theta_1,\ldots,\theta_N)_{\epsilon_1\cdots\epsilon_N}
\phantom{aaaaaa} \nonumber \\ & \times & 
^{\epsilon_N\cdots\epsilon_1}F_{a'}(\theta_N,\ldots,\theta_1)
\exp \left( - \vert {\bf r} - {\bf r}' \vert \sum_{j=1}^N
m_{\epsilon_j} \cosh \theta_j \right) . \label{3.1}  
\end{eqnarray}
Here, the first $N=0$ term of the series corresponds to the decoupling 
$\langle {\cal O}_a \rangle \langle {\cal O}_{a'} \rangle$.
The next $N\ge 1$ terms contain the form factors
\begin{eqnarray}
F_a(\theta_1,\ldots,\theta_N)_{\epsilon_1\cdots\epsilon_N}
& = & \langle 0 \vert {\cal O}_a({\bf 0}) \vert 
Z_{\epsilon_1}(\theta_1),\ldots,Z_{\epsilon_N}(\theta_N) \rangle ,
\label{3.2} \\ 
^{\epsilon_N\cdots \epsilon_1}F_{a'}(\theta_N,\ldots,\theta_1)
& = & \langle Z_{\epsilon_N}(\theta_N),\ldots,Z_{\epsilon_1}(\theta_1) 
\vert {\cal O}_{a'}({\bf 0}) \vert 0 \rangle , \label{3.3}
\end{eqnarray}
which are the matrix elements of the local operator at the origin ${\bf 0}$, 
between an $N$-particle superposition of free one-particle states, denoted as 
$\vert Z_{\epsilon}(\theta) \rangle$, and the vacuum $\vert 0 \rangle$.
The form factors, which depend on particle rapidities $\{ \theta_j \}$
only through their differences $\theta_{jk} = \theta_j - \theta_k$,
can be obtained explicitly in an axiomatic way \cite{Smirnov}.
The form-factor representation of two-point correlation functions
(\ref{3.1}) is particularly useful for large distances
$\vert {\bf r}-{\bf r}' \vert$ since it provides a systematic
large-distance expansion.
In the limit $\vert {\bf r}-{\bf r}' \vert \to \infty$, the
dominant contribution to the truncated correlation function
$\langle {\cal O}_a({\bf r}) {\cal O}_{a'}({\bf r}') \rangle_{{\rm T}}
= \langle {\cal O}_a({\bf r}) {\cal O}_{a'}({\bf r}') \rangle
- \langle {\cal O}_a \rangle \langle {\cal O}_{a'} \rangle$
comes in (\ref{3.1}) from a multi-particle intermediate state with the
minimum value of the total particle mass $\sum_{j=1}^N m_{\epsilon_j}$,
at the point of vanishing rapidities $\{ \theta_j \to 0\}$.
The corresponding exponential decay 
$\exp(-\vert {\bf r}-{\bf r}' \vert \sum_{j=1}^N m_{\epsilon_j})$
is modified by a slower (inverse power law) decaying function
which particular form depends on the form factors.
The subleading large-distance contribution to
$\langle {\cal O}_a({\bf r}) {\cal O}_{a'}({\bf r}') \rangle_{{\rm T}}$
is determined by a multiparticle state with the first ``excited''
value of the total particle mass $\sum_{j=1}^N m_{\epsilon_j}$,
and so on.

In view of the sine-Gordon representations (\ref{2.9}) and (\ref{2.10}),
the pair correlation function (\ref{2.5}) is expressible as
\begin{equation} \label{3.4}
h_{qq'}(\vert {\bf r}-{\bf r}'\vert) =
\frac{\langle {\rm e}^{{\rm i}qb\phi({\bf r})}
{\rm e}^{{\rm i}q'b\phi({\bf r}')} \rangle_{\rm T}}{\langle 
{\rm e}^{{\rm i}qb\phi} \rangle \langle {\rm e}^{{\rm i}q'b\phi} \rangle} .
\end{equation}
This means that the local operator of interest is
${\cal O}_a({\bf r}) = \exp[{\rm i}a\phi({\bf r})]$ 
with $a = q b$ $(q=\pm 1)$.

The multiparticle intermediate states of the 2D sine-Gordon model can be
created from an arbitrary number of soliton-antisoliton $S-{\bar S}$ pairs
and breathers $\{ B_j; j=1,2,\ldots<p^{-1}\}$.
From the point of view of the expansion (\ref{3.1}), the most relevant
states are those with smaller values of the total particle mass.
The general analysis of the total masses of multiparticle states 
for an arbitrary value of the temperature parameter $p<1$ is complicated.
In this paper, we are interested in the high-temperature region of relatively 
small values of the parameter $p$.
In this region, the breathers have their masses (\ref{2.13}) much
smaller than $2M$ and therefore constitute the building elements of the
most relevant multiparticle states.
In other words, the soliton-antisoliton pair of the mass $2M$ can be 
``excluded'' from the construction of the relevant multiparticle states, 
it only restricts the validity of the obtained results to specific regions 
of sufficiently small values of $p$.

In the high-temperature regime $p\to 0$, the breather masses become
integer multiples of the inverse Debye length, see formula (\ref{2.19}).
Consequently, a coalescence of the total masses of different multi-breather 
states can appear in that regime. 
In order to formalize this important phenomenon, the multi-breather states 
are grouped into subsets $\{ S_j \}_{j=1}^{\infty}$ such that a given subset 
$S_j$ contains all states possessing in the regime $p\to 0$ the same total 
particle mass equal to $j\kappa$. 
We present first few of these subsets.     
\begin{itemize}
\item
The first subset $S_1$ consists of only the one-particle $B_1$ state
with the lightest mass $m_1$. The inequality $m_1<2 M$ holds in 
the whole stability interval $p<1$ ($\beta<2$). 
\item
The second subset $S_2$ consists of the one-particle $B_2$ state of
the mass $m_2$ and the two-particle $B_1B_1$ state of the mass $2 m_1$. 
The inequalities $m_2\le 2m_1<2M$ hold in the interval $p<1/3$ ($\beta<1$).
The equality $m_2=2m_1$ takes place in the limit $p\to 0$.
\item
The third subset $S_3$ consists of the one-particle $B_3$ state of
the mass $m_3$, the two-particle $B_1B_2$ state of the mass $m_1+m_2$
and the three-particle $B_1B_1B_1$ state of the mass $3m_1$. 
The inequalities $m_3\le m_1+m_2\le 3m_1<2M$ hold 
for $p<(2/\pi)\arcsin(1/3)\sim 0.216$.
The equalities $m_3=m_1+m_2=3m_1$ take place in the limit $p\to 0$.
\end{itemize}
From the point of view of the next analysis, only the first two
subsets are relevant.
They contain the one-particle $B_1$ state, the one-particle $B_2$ state and
the two-particle $B_1B_1$ state.
The explicit form factors for these states are summarized 
in the next two paragraphs.

For the $B_1$ breather, the multi-breather form factors
$F_a(\theta_1,\ldots,\theta_N)_{1\ldots 1}$ and
$^{1\ldots 1}F_{a'}(\theta_N,\ldots,\theta_1)$ 
$=$ $F_{a'}(\theta_N,\ldots,\theta_1)_{1\ldots 1}$
are presented for the exponential operator 
${\cal O}_a({\bf r}) = \exp[{\rm i}a\phi({\bf r})]$ in ref. \cite{Lukyanov}.
Within the notation (\ref{3.2}), in the special case of interest $a=q b$ 
$(q=\pm 1)$ they read
\begin{eqnarray}
\langle 0 \vert {\rm e}^{{\rm i}qb\phi} \vert B_1(\theta) \rangle
& = & - {\rm i} q (\pi \lambda)^{1/2} \langle {\rm e}^{{\rm i}qb\phi} \rangle ,
\label{3.5} \\
\langle 0 \vert {\rm e}^{{\rm i}qb\phi} \vert B_1(\theta_2),
B_1(\theta_1) \rangle 
& = & - (\pi \lambda) R(\theta_1 - \theta_2) 
\langle {\rm e}^{{\rm i}qb\phi} \rangle , \label{3.6}
\end{eqnarray}
etc.
Here, the parameter $\lambda$ is defined by
\begin{equation} \label{3.7}
\lambda = \frac{4}{\pi} \sin(p\pi) \cos \left( \frac{p\pi}{2}
\right) \exp \left( - \int_0^{p\pi} \frac{{\rm d}t}{\pi}
\frac{t}{\sin t} \right) 
\end{equation}
and the function $R(\theta)$ is given on the interval
$-2\pi + p \pi < {\rm Im}(\theta) < - p \pi$ by
\begin{eqnarray} 
R(\theta) & = & {\cal N} \exp \Bigg\{ 8 \int_0^{\infty}
\frac{{\rm d}t}{t} \frac{\sinh(t) \sinh(p t) \sinh[(1+p)t]}{\sinh^2(2t)}
\nonumber \\ & & \quad \times
\sinh^2 \left[ t\left( 1 - \frac{{\rm i}\theta}{\pi} \right) \right] \Bigg\} , 
\label{3.8} \\ {\cal N} & = & \exp \left\{ 4 \int_0^{\infty}
\frac{{\rm d}t}{t} \frac{\sinh(t) \sinh(p t) \sinh((1+p)t)}{\sinh^2(2t)} 
\right\} . \label{3.9}
\end{eqnarray}
The function $R(\theta)$ satisfies two useful relations:
\begin{equation} \label{3.10}
R(\theta) R(\theta \pm {\rm i}\pi) =
\frac{\sinh(\theta)}{\sinh(\theta) \mp {\rm i} \sin(p \pi)} 
\end{equation}
and
\begin{equation} \label{3.11}
R(-\theta) = S_{11}(\theta) R(\theta) ,
\end{equation}
where
\begin{equation} \label{3.12}
S_{11}(\theta) = 
\frac{\sinh(\theta)+{\rm i}\sin(p\pi)}{\sinh(\theta)-{\rm i}\sin(p\pi)}
\end{equation}
is the $B_1B_1$ scattering matrix \cite{Zamolodchikov79}.
These relations enable one to extend the definition (\ref{3.8})-(\ref{3.9})
of $R(\theta)$ to arbitrary values of ${\rm Im}(\theta)$. 
For the case of special interest ${\rm Im}(\theta)=0$,
one gets the representation
\begin{eqnarray} 
R(\theta) & = & \frac{\sinh(\theta)}{\sinh(\theta)+{\rm i}\sin(p\pi)}
\label{3.13} \\ & & \times
\exp \Bigg\{ - 4 \int_0^{\infty} \frac{{\rm d}t}{t} 
\frac{\sinh(t) \sinh(p t) \sinh[(1+p)t]}{\sinh^2(2t)}
\cos\left( \frac{2\theta t}{\pi}\right) \Bigg\} \nonumber
\end{eqnarray}
valid in the whole stability region $0\le p<1$.

The $B_2$ breather is a boundstate of the two $B_1$ breathers
since the $B_1B_1$ scattering matrix (\ref{3.12}) has the $B_2$ pole
at $\theta={\rm i}p\pi$.
Consequently, the one-particle $B_2$ form factor can be calculated from 
the two-particle $B_1B_1$ form factor (\ref{3.6}) using 
a bootstrap procedure \cite{Smirnov,Mussardo}:
\begin{equation} \label{3.14}
\Gamma \langle 0\vert {\rm e}^{{\rm i}qb\phi} \vert B_2(\theta) \rangle
= {\rm i}\, {\rm res}_{\epsilon=0} \left\{ - (\pi\lambda) 
R(-{\rm i}p\pi+\epsilon) \langle {\rm e}^{{\rm i}qb\phi} \rangle \right\} ,
\end{equation}
where the parameter $\Gamma$ is related to the residue of the $B_2$ pole
in the $B_1B_1$ scattering as follows
\begin{equation} \label{3.15}
\Gamma = \left[ - {\rm i}\, {\rm res}_{\theta={\rm i}p\pi} S_{11}(\theta) 
\right]^{1/2} = \sqrt{2\tan(p\pi)} .
\end{equation}
Using the relation (\ref{3.10}) for $\theta=-{\rm i}p\pi+\epsilon$
($\epsilon\to 0$), one finds that
\begin{equation} \label{3.16}
\langle 0 \vert {\rm e}^{{\rm i}qb\phi} \vert B_2(\theta) \rangle
= - (\pi\lambda) \left[ \frac{\tan(p\pi)}{2} \right]^{1/2}
\frac{1}{R[-{\rm i}\pi(1+p)]} \langle {\rm e}^{{\rm i}qb\phi} \rangle .
\end{equation}

Using the explicit form factors (\ref{3.5}), (\ref{3.6}) and (\ref{3.16})
in the series representation (\ref{3.1}) for the pair correlation
function (\ref{3.4}), one finally arrives at
\begin{eqnarray}
h_{qq'}(r) & = & - q q' \lambda K_0(m_1 r)
+ \pi \lambda^2 \frac{\tan(p\pi)}{2} \frac{1}{R^2[-{\rm i}\pi(1+p)]}
K_0(m_2 r) \nonumber \\
& & + \frac{\lambda^2}{2!} I(m_1 r) + o\left({\rm e}^{-m_3 r}\right) ,
\label{3.17}
\end{eqnarray}
where 
\begin{equation} \label{3.18}
K_0(x) = \int_{-\infty}^{\infty} \frac{{\rm d}\theta}{2}\,
{\rm e}^{-x\cosh\theta}
\end{equation}
is the modified Bessel function of second kind \cite{Gradshteyn}
and $I(m_1 r)$ denotes the double integral
\begin{equation} \label{3.19}
I(m_1 r) = \int_{-\infty}^{\infty} \frac{{\rm d}\theta_1}{2}
\int_{-\infty}^{\infty} \frac{{\rm d}\theta_2}{2}
R(\theta_1-\theta_2) R(\theta_2-\theta_1)\,
{\rm e}^{-m_1 r(\cosh\theta_1+\cosh\theta_2)} .
\end{equation}
Based on the previous mass analysis for relevant multiparticle states, 
the asymptotic expansion (\ref{3.17}) as a whole applies in the region 
$0\le p<1/3$ ($0\le \beta<1$).

\renewcommand{\theequation}{4.\arabic{equation}}
\setcounter{equation}{0}

\section{Asymptotic behavior of pair correlations}

\subsection{Charge correlation function}
According to the definition of the charge correlation function $h_{\rho}$ 
in (\ref{2.6}), only the first term on the right-hand side (rhs) of
Eq. (\ref{3.17}), proportional to $q q'$, contributes to $h_{\rho}$.
Since the modified Bessel function $K_0(x)$ has the asymptotic form
\begin{equation} \label{4.1}
K_0(x) \mathop{\sim}_{x\to\infty}
\left( \frac{\pi}{2 x} \right)^{1/2} {\rm e}^{-x} ,
\end{equation}
the large-distance behavior of $h_{\rho}$, valid in the whole stability
region $0\le\beta<2$, reads
\begin{equation} \label{4.2}
h_{\rho}(r) \mathop{\sim}_{r\to\infty} - \lambda
\left( \frac{\pi}{2 m_1 r} \right)^{1/2} {\rm e}^{-m_1 r} .
\end{equation}
This formula tells us that the mass $m_1$ of the lightest $B_1$ breather is 
the renormalized inverse screening length of the charge correlation function.

The mass $m_1$, given by Eq. (\ref{2.18}), 
has the small-$\beta$ expansion of the form
\begin{eqnarray} 
m_1 & = & \kappa \left[
\frac{\sin(\pi\beta/(4-\beta))}{\pi\beta/(4-\beta)} \right]^{1/2}
\nonumber \\
& = & \kappa \left[ 1 - \frac{\pi^2}{192} \beta^2
- \frac{\pi^2}{384} \beta^3 + O(\beta^4) \right] \label{4.3}
\end{eqnarray}
and the parameter $\lambda$, given by Eq. (\ref{3.7}), 
has the small-$\beta$ expansion of the form
\begin{equation} \label{4.4}
\lambda = \beta \left[ 1 - \left( \frac{1}{32} + \frac{7\pi^2}{384} \right)
\beta^2 - \left( \frac{1}{96} + \frac{23\pi^2}{2304} \right) \beta^3
+ O(\beta^4) \right] .
\end{equation}
When $\beta\to 0$, one has $m_1\sim\kappa$ and $\lambda\sim\beta$,
so that the formula (\ref{4.2}) reproduces correctly the DH result
(\ref{A.9}).
We conclude that the large-distance behavior of the charge correlation
function changes continuously when going from strictly positive values of 
$\beta>0$ to the regime $\beta\to 0$, in agreement with the general belief.

\subsection{Number density correlation function}
According to the definition of the number density correlation function
$h_n$ in (\ref{2.6}), the second and third terms on the rhs of 
Eq. (\ref{3.17}), which do not depend on the signs of the considered
charges $q$ and $q'$, contribute to $h_n$:
\begin{equation} \label{4.5}
h_n(r) =  \pi \lambda^2 \frac{\tan(p\pi)}{2} \frac{1}{R^2[-{\rm i}\pi(1+p)]}
K_0(m_2 r) + \frac{\lambda^2}{2!} I(m_1 r) + o\left({\rm e}^{-m_3 r}\right) .
\end{equation}
This asymptotic formula holds in the region $0\le p<1/3$ ($0\le\beta<1$).

The exponential decay at large $r$ of the two terms in Eq. (\ref{4.5})
is given by the asymptotic of $K_0(m_2 r)\sim \exp(-m_2 r)$ 
and $I(m_1 r)\sim \exp(-2m_1 r)$.
At strictly positive $\beta>0$, the inequality 
$m_2=2 m_1 \cos(p\pi/2) < 2m_1$ takes place and
therefore the large-distance behavior of $h_n(r)$ is dominated by
\begin{equation} \label{4.6}
h_n(r) \mathop{\sim}_{r\to\infty} \pi \lambda^2 \frac{\tan(p\pi)}{2} 
\frac{1}{R^2[-{\rm i}\pi(1+p)]} \left( \frac{\pi}{2 m_2 r} \right)^{1/2}
{\rm e}^{-m_2 r} .
\end{equation}
As concerns the $\beta\to 0$ limit of this formula, 
considering $\lambda\to\beta$, $p\to\beta/4$, $R(-{\rm i}\pi)\to 1$ 
and $m_2\to 2\kappa$ leads to the expression
\begin{equation} \label{4.7}
h_n(r) \mathop{\sim}_{r\to\infty} \frac{\pi^2\beta^3}{8} 
\left( \frac{\pi}{4\kappa r} \right)^{1/2} {\rm e}^{-2\kappa r} .
\end{equation}

The large-distance form of the density correlation function is
derived at lower orders in $\beta$ by a systematic diagrammatic
expansion in Appendix, see formula (\ref{A.18}).
It is seen that the leading high-temperature term is of order
$\beta^2$, but when one is interested in the large-distance
asymptotics $\kappa r\to\infty$, the leading term is of order $\beta^3$. 
This term is twice smaller than the obtained result (\ref{4.7}). 
The reason for this inconsistency consists in the fact that in the
regime $\beta\to 0$ the coalescence of the inverse correlation lengths 
$m_2=2 m_1\to 2\kappa$ takes place in Eq. (\ref{4.5}).
As a consequence, also the $B_1B_1$ term $(\lambda^2/2!) I(m_1 r)$
in (\ref{4.5}), which is subleading for strictly positive $\beta>0$,
contributes to the large-$r$ asymptotic behavior of $h_n(r)$ 
when $\beta\to 0$.
The derivation of this additional contribution and the subsequent
verification of the consistency of the final result with 
the high-temperature asymptotic formula (\ref{A.18}) are the subjects 
of the next paragraph.

To evaluate the $\beta\to 0$ (or, equivalently, $p\to 0$) behavior of 
the integral $I(m_1 r)$ given by (\ref{3.19}), we first use 
the formula (\ref{3.13}), valid for real values of $\theta$, to write down
\begin{equation} \label{4.8}
R(\theta) R(-\theta) \mathop{\sim}_{p\to 0} 
\frac{\sinh^2(\theta)}{\sinh^2(\theta)+\sin^2(p\pi)}
= 1 - \frac{\sin^2(p\pi)}{\cosh^2(\theta) - \cos^2(p\pi)} .
\end{equation}
The change of variables $\theta_t=(\theta_1+\theta_2)/2$ and 
$\theta=\theta_1-\theta_2$ in the integral (\ref{3.19}) and the subsequent 
integration over $\theta_t$ then leads to the representation
\begin{equation} \label{4.9}
I(m_1 r) \mathop{\sim}_{p\to 0} 
K_0^2(m_1 r) - \int_{-\infty}^{\infty} \frac{{\rm d}\theta}{2}
\frac{\sin^2(p\pi)}{\cosh^2(\theta) - \cos^2(p\pi)} 
K_0\left[ 2 m_1 r \cosh(\theta/2)\right] .
\end{equation} 
It is easy to show that for any function $f(\theta)$ regular
at $\theta=0$ it holds
\begin{equation} \label{4.10}
\int_{-\infty}^{\infty} \frac{{\rm d}\theta}{2}
\frac{1}{\cosh(\theta)-\cos(p\pi)} f(\theta)
\mathop{\sim}_{p\to 0} \frac{1}{p} f(0) .
\end{equation}
The application of this relation to the integral in Eq. (\ref{4.9}) leads to
\begin{equation} \label{4.11}
I(m_1 r) \mathop{\sim}_{p\to 0} K_0^2(m_1 r) - 
\frac{p\pi^2}{2} K_0(2 m_1 r) .
\end{equation}
Thus,
\begin{equation} \label{4.12}
\frac{\lambda^2}{2!} I(m_1 r) = \frac{\beta^2}{2} K_0^2(\kappa r)
- \frac{\pi^2\beta^3}{16} K_0(2\kappa r) + O(\beta^4) .
\end{equation}
At asymptotically large distance,
\begin{equation} \label{4.13}
\frac{\lambda^2}{2!} I(m_1 r) \mathop{\sim}_{r\to\infty}
\frac{\pi\beta^2}{4\kappa r} {\rm e}^{-2\kappa r} - \frac{\pi^2\beta^3}{16} 
\left( \frac{\pi}{4\kappa r} \right)^{1/2} {\rm e}^{-2\kappa r}
+ O(\beta^4) .
\end{equation}
Summing up the rhs of this formula with the rhs of the previous formula 
(\ref{4.7}), one recovers correctly the asymptotic result (\ref{A.18})
obtained by the systematic $\beta$ expansion. 
We conclude that the large-distance behavior of the number density
correlation function undertakes an abrupt change, namely the discontinuity,
when going from strictly positive values of $\beta>0$ to the $\beta\to 0$
regime.

We have shown by the exact calculation that the large-distance asymptotics
of the density correlation function at fixed temperature $\beta>0$
does not coincide with that obtained when the high-temperature 
$\beta$-expansion has been performed first.
This is equivalent to saying that the large-distance asymptotics and 
the high-temperature limit do not commute for this function.
Such phenomenon is in contradiction with the ``usual'' physical intuition.

\renewcommand{\theequation}{5.\arabic{equation}}
\setcounter{equation}{0}

\section{Conclusion}
In the present paper, we took advantage of the exact solvability 
of the bulk 2D Coulomb gas to study the large-distance behavior of 
correlation functions between charged particles.
Using the form-factor technique for the equivalent (1+1)-dimensional
sine-Gordon theory, we have expressed in Eqs. (\ref{3.17})-(\ref{3.19})
the leading and subleading asymptotic terms of particle correlation 
functions in terms of the masses of breathers belonging 
to the sine-Gordon particle spectrum.

The result for the charge correlation function $h_{\rho}(r)$
(\ref{4.2})-(\ref{4.4}), valid in the whole stability region 
of the Coulomb gas $0<\beta<2$, has the generally anticipated property: 
in the $\beta\to 0$ regime, it reduces continuously to 
the DH result (\ref{A.9}).
This means that heuristic extensions of mean-field theories
to finite temperatures \cite{Levin} are reasonable when they are
based on the charge-charge correlations.

On the other hand, the formula (\ref{4.6}) for the asymptotic decay
of the number density correlation function $h_n(r)$, valid in the region 
$0<\beta<4/3$, does not reproduce in the $\beta\to 0$ regime 
[see Eq. (\ref{4.7})] the result of the high-temperature expansion
(\ref{A.18}).
The reason for this inconsistency consists in the fact that when
$\beta\to 0$ the term $(\lambda^2/2!)I(m_1 r)$ in Eq. (\ref{4.5}),
which is subleading for strictly positive $\beta>0$, interferes 
with the leading one and also contributes to the asymptotic result.
Taking into account the asymptotic formula (\ref{4.13}) for this
subleading term, one recovers correctly the high-temperature formula
(\ref{A.18}).
As a consequence of the above scenario, the large-distance 
behavior of the number density correlation function undertakes 
a discontinuity when going from strictly positive values of $\beta>0$ 
to the $\beta\to 0$ regime.
The high-temperature expansion (\ref{A.18}) therefore does not reflect 
adequately the large-distance behavior of the number density correlation 
at strictly positive $\beta>0$.
This phenomenon contradicts the general belief and one has to be very
careful when extending the DH description of number density correlations
to finite temperatures.

We notice that the derivation of the standard Debye-H\"uckel theory 
is based on electrical quantities and it is perhaps not surprising that the
behavior of quantities related to the number density is not always
adequately reproduced in this high-temperature theory.

The anomaly in the large-distance behavior of the number density correlation
function $h_n(r)$ was observed due to the availability of the exact 
(and, therefore, {\em nonperturbative}) description of the 2D Coulomb gas.
The anomaly could be observed perturbatively only after the resummation 
of specific diagrammatic contributions in all $\beta$ orders of 
the large-distance decay of $h_n(r)$.
It would be interesting to reveal the resummation mechanism because
the described anomaly might be present in 3D Coulomb fluids, too.

\renewcommand{\theequation}{A.\arabic{equation}}
\setcounter{equation}{0}

\section*{Appendix: Diagrammatic expansion}
In this appendix, we derive the asymptotic large-distance form
of the charge and number density correlation functions in the bulk 2D
Coulomb gas, at lower orders in $\beta$.
The $\beta$ expansions of the correlation functions must be taken 
for a fixed value of the inverse Debye length $\kappa=\sqrt{2\pi\beta n}$, 
which only sets the length scale.

For the considered Coulomb gas with the charge symmetry, 
the ordinary Ornstein-Zernike (OZ) equation splits into two independent 
relations for the charge and density functions \cite{Jancovici00}
\begin{eqnarray} 
h_{\rho} & = & c_{\rho} + c_{\rho} \ast n \ast h_{\rho} , \label{A.1} \\
h_n & = & c_n + c_n \ast n \ast h_n , \label{A.2}
\end{eqnarray}
where $\ast$ denotes a convolution product and the charge and density
{\em direct} correlation functions are defined in analogy with 
Eq. (\ref{2.6}) as follows
\begin{equation} \label{A.3}
c_{\rho} = \frac{1}{4} \sum_{q,q'=\pm 1} q q' c_{qq'} , \quad
c_n = \frac{1}{4} \sum_{q,q'=\pm 1} c_{qq'} .
\end{equation}

In the renormalized Mayer expansion of the excess Helmholtz free energy
in density \cite{Jancovici00}, the chains of simple $-\beta v$ bonds
are summed up into the renormalized bonds $K$, defined implicitly 
by the relation
\begin{equation} \label{A.4}
K =  -\beta v + (-\beta v)\ast n \ast K 
\end{equation}
with $v$ being the Coulomb potential.
In the infinite 2D space, one has
\begin{equation} \label{A.5} 
K(r) = - \beta K_0(\kappa r) ,
\end{equation}
where $K_0$ is the modified Bessel function of second kind (\ref{3.18}).

In the renormalized-bond format, the charge direct correlation function 
is expressible as
\begin{equation} \label{A.6}
c_{\rho}(r) = - \beta v(r) + 
\sum_{j=3}^{\infty} \beta^j c_{\rho}^{(j)}(r) ,
\end{equation}
where only such renormalized graphs of the excess Helmholtz free energy 
contribute to the coefficients $\{ c_{\rho}^{(j)} \}_{j=3}^{\infty}$ 
which have their two root vertices with an odd bond-coordination 
and their field vertices with an even bond-coordination.
In particular \cite{Samaj02a},
\begin{eqnarray}
c_{\rho}^{(3)}(r) & = & - \frac{1}{6} K_0^3(\kappa r) , \label{A.7} \\
c_{\rho}^{(4)}(r) & = & - \frac{1}{8\pi} K_0(\kappa r)
\int {\rm d}^2 (\kappa r') K_0^2(\kappa r')
K_0^2(\kappa\vert {\bf r}-{\bf r}'\vert) , \label{A.8}
\end{eqnarray}
etc.
Inserting the leading $c_{\rho}(r) = -\beta v(r)$ into 
the OZ relation (\ref{A.1}), one gets for $h_{\rho}$ nothing but 
the definition (\ref{A.4}) of the renormalized bond $K$.
Therefore, at the lowest order in $\beta$,
\begin{equation} \label{A.9}
h_{\rho}(r) = - \beta K_0(\kappa r) \mathop{\sim}_{r\to\infty}
- \beta \left( \frac{\pi}{2\kappa r} \right)^{1/2}
\exp ( - \kappa r )
\end{equation}
since $K_0(x)$ has the asymptotic form (\ref{4.1}).
It was shown in Ref. \cite{Samaj02a} that the consideration of the terms 
$j=3,4$ in the expansion (\ref{A.6}) with the corresponding coefficients 
(\ref{A.7}) and (\ref{A.8}) implies the large-distance asymptotic behavior 
(\ref{4.2}) with the $\beta$ expansions of the parameters $m_1$ (\ref{4.3}) 
and $\lambda$ (\ref{4.4}) up to the indicated $\beta^3$ order.

Within the renormalized-bond formalism, the number density direct correlation 
function is given by
\begin{equation} \label{A.10}
c_n(r) =  \frac{1}{2!} K^2(r) + \sum_{j=4}^{\infty} \beta^j c_n^{(j)}(r) .
\end{equation}
Here, the leading term of order $\beta^2$ corresponds to 
the renormalized Meeron (watermelon) diagram and renormalized graphs 
of the excess Helmholtz free energy, contributing to the coefficients 
$\{ c_n^{(j)} \}_{j=4}^{\infty}$, are the ones which have their two root 
vertices as well as field vertices with an even bond-coordination.
Inserting the leading $c_n(r) =  \frac{1}{2!} K^2(r)$ of order $\beta^2$
into the OZ relation (\ref{A.2}), the convolution term is easily seen 
to be of higher order $\beta^3$. 
Thus, at lowest order in $\beta$, $h_n(r) = c_n(r)$, 
\begin{equation} \label{A.11}
h_n(r) = \frac{\beta^2}{2} K_0^2(\kappa r) \mathop{\sim}_{r\to\infty}
\frac{\pi\beta^2}{4\kappa r} \exp (-2 \kappa r) .
\end{equation}
Since the sum on the rhs of (\ref{A.10}) starts from $j=4$, the
$\beta^3$ term of $h_n$ has its origin exclusively in the convolution
term of the OZ relation (\ref{A.2}) taken with $c_n(r) = h_n(r) = K^2(r)/2!$. 
Consequently,
\begin{equation} \label{A.12}
h_n(r) = \frac{\beta^2}{2} K_0^2(\kappa r) 
+ \frac{\beta^3}{4} J(\kappa r) + O(\beta^4) ,
\end{equation}
where the integral $J$ is defined by
\begin{equation} \label{A.13}
J(r) = \int \frac{{\rm d}^2 r'}{2\pi} K_0^2({\bf r}') K_0^2({\bf r}-{\bf r}') .
\end{equation} 
In terms of the 2D Fourier transform of $K_0^2({\bf r})$
\begin{equation} \label{A.14}
G({\bf k}) = \int \frac{{\rm d}^2 r}{2\pi} 
{\rm e}^{-{\rm i}{\bf k}\cdot{\bf r}} K_0^2({\bf r})
= \frac{\ln\left[ (k/2) + \sqrt{1+(k/2)^2} \right]}{k\sqrt{1+(k/2)^2}} ,
\end{equation}
$J$ is expressible as follows
\begin{equation} \label{A.15}
J(r) = \int \frac{{\rm d}^2 k}{2\pi} {\rm e}^{{\rm i}{\bf k}\cdot{\bf r}} 
G^2(k) .
\end{equation}
Let us put ${\bf r} = (0,r)$ in the Cartesian notation. 
Since the function $G^2(k)$ has simple poles at 
$k_y = \pm {\rm i}\sqrt{4+k_x^2}$, the integration over the vector
component $k_y$ can be performed explicitly as the contour integration
in the complex plane by using the residuum theorem, with the result
\begin{equation} \label{A.16}
J(r) = \frac{\pi^2}{4} \int_{-\infty}^{\infty} \frac{{\rm d}k_x}{2}
\frac{1}{\sqrt{4+k_x^2}} \exp\left( - r \sqrt{4+k^2} \right) .
\end{equation}
The integral in (\ref{A.16}) is equal to $K_0(2r)$ and so
$J(r) = (\pi^2/4) K_0(2r)$.
Eq. (\ref{A.12}) thus takes the form
\begin{equation} \label{A.17} 
h_n(r) = \frac{\beta^2}{2} K_0^2(\kappa r) 
+ \frac{\pi^2\beta^3}{16} K_0(2\kappa r) + O(\beta^4) .
\end{equation}
At asymptotically large distance,
\begin{equation} \label{A.18}
h_n(r) \mathop{\sim}_{r\to\infty}
\frac{\pi\beta^2}{4\kappa r} {\rm e}^{-2\kappa r}
+ \frac{\pi^2\beta^3}{16} 
\left( \frac{\pi}{4\kappa r} \right)^{1/2} {\rm e}^{-2\kappa r}
+ O(\beta^4) .
\end{equation}
Here, the leading high-temperature term is of order $\beta^2$.
But when one is interested in the large-distance $\kappa r\to\infty$
asymptotics of the density correlation function, like in the present
paper, the leading term is of order $\beta^3$.

\section*{Acknowledgments}
I thank Bernard Jancovici for careful reading of the manuscript
and useful comments.
The support by grant VEGA 2/6071/26 is acknowledged.

\end{document}